\documentclass[aps,prb,showpacs,twocolumn,superscriptaddress,floatfix]{revtex4}
\usepackage{epsfig}
\usepackage{graphics}
\usepackage{bm}

\begin{document}
\title{ Reexamining the Einstein-Podolsky-Rosen experiment,
photon correlation and Bell's inequality}
\author{L.~Fritsche}
\thanks{Corresponding author}
\email{lfritsche@t-online.de} \affiliation{
Institut f\"ur Theoretische Physik der Technischen Universit\"at
Clausthal, D-38678 Clausthal-Zellerfeld, Germany}
\author{M.~Haugk} \affiliation{HPC-Auto, Hewlett-Packard GmbH,
D-71065 Sindelfingen, Germany}
\begin{abstract}
The purpose of this article is to show that the introduction of hidden variables to describe individual events is 
fully consistent with the statistical predictions of quantum theory. We illustrate the validity of this assertion by 
discussing two fundamental experiments on correlated photons which are believed to behave ``violently 
non-classical''. Our considerations carry over to correlated pairs of neutral particles of spin one-half in a 
singlet state. Much in the spirit of Einstein's conviction we come to the conclusion that the state vector of a 
system does not provide an exhaustive description of the individual physical system. We also briefly discuss an 
experiment on ``quantum teleportation'' and demonstrate that our completely local approach leads to a full 
understanding of the experiment indicating the absence of any teleportation phenomenon. We caution that the 
indiscriminate use of the term ``Quantum Theory'' tends to obscure distinct differences between the quantum 
mechanics of massive particles and the propagation of photons. It is emphasized that the properties of polarizers, 
beam splitters, halfwave plates etc. used in photon-correlation experiments are defined by the laws of classical 
optics. Hence, understanding the outcome of those experiments requires a well-founded interconnection between 
classical and quantum electrodynamics which we scrutinize for critical assumptions.
\end{abstract}
\pacs{03.65.Ud, 32.80.Wr, 42.25.Ja}

\maketitle

\section{Introduction}

The numerous studies on entangled states and in particular on correlated photons during the past 20 years have 
repeatedly stirred up renewed interest in the seminal paper by Einstein, Podolsky and Rosen \cite{Einstein1} 
(commonly referred to as EPR) on the issue of whether or not physical reality can exhaustively be described within 
the framework of quantum mechanics. Their analysis is based on a meanwhile popular thought experiment that deals 
with a pair of correlated massive particles. In 1951 Bohm \cite{Bohm} extended this thought experiment by 
considering two particles with spin in a singlet state. It is this form of the thought experiment that has been 
discussed in the literature ever since and will henceforth be referred to as EPRB-experiment.
\\[0.3cm]
Despite the plethora of articles and monographs that are fully or in part devoted to the EPRB-problem, there has 
been no successful attempt so far to explain the experiments within a local theory by introducing and averaging over 
a ``hidden parameter'' that represents an indispensable classical variable. We shall demonstrate that our na\"ive 
realistic local  approach does not only yield the ``correct'' quantum mechanical result but that it applies as well 
to the fundamental experiment by Aspect et al. \cite{Aspect} on correlated pairs of photons. Also in this case the 
result of our calculation is in complete agreement with the experiment.
\\[0.3cm]
The ERPB-experiment is commonly discussed in terms of pairs of particles that are emitted from a source in opposite 
directions having opposite transverse spin directions. The latter are thought to be measured by two identical  
instruments on either side of the source but sufficiently far away from it so that the respective particle passing 
through one of the measurement instrument cannot interact with the other particle running through the instrument in 
the opposite direction. Our analysis takes the na\"ive realist's standpoint that the process of measurement on 
either side of the setup is strictly local in the sense that the measurement on one side has no effect on the 
measurement on the other side. Hence we deny the possibility of what Einstein termed \textit{spooky action at a 
distance}. This implies that the removal of the measurement instrument on one side would not affect the measurement 
(in practice: the count rate) on the other side. But the original opposite orientation of the particle spins, 
dictated by the process of their generation, remains unaffected over any conceivable distance up to the entrance 
slits of the measurement instruments. This property is commonly referred to as ``perfect correlation''. We question 
the validity of the standard assumption that the plane of polarization spanned by the two spins and the line of 
particle propagation is unknown and therefore indetermined in advance of measurement. Instead we identify the angle 
enclosed by the normal of this plane and some laboratory-fixed axis as a ``hidden parameter'' which attains the 
character of a random variable as one repeats the generation of pairs a sufficiently large number of times.
\\[0.3cm]
In Section \ref{correlated_pairs} we shall first discuss the case of ``entangled'' photons as the respective 
experiment
 has actually been carried out with considerable sophistication and great care \cite {Aspect}. This applies as well 
to many similar experiments on correlated photons, one of which \cite{Weinfurter} will be the subject of Section 
\ref{Multi-photon}. By contrast, experiments on correlated pairs of massive particles and opposite transverse spin, 
which we discuss in Sections
\ref{Bell-message} and \ref{Locality}, are mostly fictitious or less complete.
\section{ Correlated pairs of photons}\label{correlated_pairs}
\textit{All the years of willful pondering have not brought me any closer to the an\-swer to the question ``what are 
light quanta''. Today every good-for-nothing believes he should know it, but he is mistaken...}
\\[0.2cm]
\textbf{Albert Einstein}\\
(In a letter to M. Besso, 1951)
\\[0.2cm]
Since a photon (or ``light quantum'') that has been emitted, for example, from an excited hydrogen atom delivers its 
energy completely to an absorber hydrogen atom, even when this atom is at an astronomical distance, we picture a 
photon as a point-like particle which is indivisible. This will also prove to be consistent with its properties 
displayed in beam splitters and polarizers.  As one knows, for example, from Schr\"odinger's theory of the Doppler 
shift of atomic radiation \cite {Schrodinger} the photon's recoil is transferred to the emitter once it has been 
ejected, regardless whether or not it is absorbed some time later. Hence the assertion that a photon comes into 
existence only when it is observed (or ``measured'') appears to have little justification.
\\[0.2cm]
In addition, there is a widespread belief that a photon cannot be associated with a certain polarization unless this 
property has been measured. We advance the opinion that this assertion is as implausible as unjustified. The 
indeterminacy of photon polarization before its measurement represents one of those arbitrary and unnecessary 
assumptions that lead inescapably to invoking spooky action at a distance. \\
A crucial component of all the measurement instruments dealing with photon experiments consists in polarizers into 
which the photons under study penetrate. It is rarely discussed in the meanwhile enormous literature on this subject 
that the incoming photon is absorbed within the polarizer after a few wavelengths and coherently replaced by another 
one whose polarization lies in one of the two orthogonal planes of the uniaxial birefringent material. One of the 
planes is spanned by the optic axis of the material and the normal of the face of incidence. The absorption of 
photons after a short distance of travelling in an optical material has most strikingly been demonstrated by Beth 
\cite{Beth} who used a quarter wave plate to absorb linearly polarized photons and to convert them into circularly 
polarized photons whose angular momenta cause a circular recoil momentum in the plate. The plate was mounted 
horizontally at a vertical quartz fiber and responded to the angular momentum transfer by twisting the fiber to a 
certain maximum angle. Using this technique in a torsional pendulum device Beth could determine the angular momentum 
of circularly polarized photons.\\
The disappearance of the incoming wave in favor of a secondary wave is the content of the fundamental 
\textit{Ewald-Oseen} extinction theorem. (See, for example: M.~Born and E.~Wolf \cite{Born_Wolf}.)
As soon as a wave train (associated with a photon) penetrates into an optical material it excites coherently its 
atoms and causes them to set up a secondary wave field. Thereby the wave train loses its energy. For simplicity we 
disregard in the following the slight departure from monochromacy in going from a plane wave to a wave packet and 
characterize the photon state of the interaction-free wave train by $|n_{\vec{k}}\rangle$ where $\vec{k}$ denotes 
its wave vector. Its state at finite coupling to the atoms of the optical material may be described by
\begin{eqnarray}
\label{eqn:photonstate}
|\phi_{\gamma}(t)\rangle=c_1(t)\,|n_{\vec{k}}^{(1)}\rangle+c_2(t)\,|n_{\vec{k}}^{(2)}\rangle\,,
\end{eqnarray}
where
$$
n_{\vec{k}}^{(1)}=1 \quad \mbox{and}\quad n_{\vec{k}}^{(2)}=0 \:\mbox{(for the vacuum state)}\,.
$$
We assume a simple plausible time-dependence of the coefficients
\begin{eqnarray*}
c_1(t)=\frac{1}{\sqrt{2}}\,[1-\tanh(2\,t/\tau)]^{\frac{1}{2}}\\ 
c_2(t)=\frac{1}{\sqrt{2}}\,[1+\tanh(2\,t/\tau)]^{\frac{1}{2}}
\end{eqnarray*}
which have the property
$$
c_1^2(t)+c_2^2(t)=1
$$
and hence ensure the norm unity of $|\phi_{\gamma}\rangle$. The quantity $\tau$ denotes the absorption time. We 
reference the middle of the absorption time interval to $t=0$.\\
Since we have for simplicity substituted the wave train by a plane wave, the operator $\hat{E}$ of the electric 
field may be reduced to one term
\begin{eqnarray*}
\hat{E}(\vec{r},t)=\sqrt 
{\frac{\hbar\,\omega_{\vec{k}}}{2\,\varepsilon_0\,V}}\;\vec{\epsilon}_{\vec{k}}\,\left[\hat{a}_{\vec{k}}\,e^{i(\vec 
{k}\cdot \vec{r}-\omega_{\vec{k}}\,t)}+\hat{a}_{\vec{k}}^{\dagger}\,e^{-i(\vec {k}\cdot 
\vec{r}-\omega_{\vec{k}}\,t)}\right]\,,
\end{eqnarray*}
where $V$ is the normalization volume, $\omega_{\vec{k}}$ the frequency, 
$\hat{a}_{\vec{k}}^{\dagger}\;,\hat{a}_{\vec{k}}$ denote the photon creation and annihilation operator,  
$\vec{\epsilon}_{\vec{k}}$ is the unit vector of polarization, and $\epsilon_0$ denotes the vacuum permittivity. We 
have, further, introduced Planck's constant $h$ in the form $\hbar=h/2\,\pi$. If one uses the commutation rules for 
$\hat{a}_{\vec{k}}^{\dagger}$ and $\hat{a}_{\vec{k}}$
it is straight-forward to show that the expectation value of $\hat{E}$ can be cast as
\begin{eqnarray}
\label{eqn:oscillating_E}
\vec{E}(\vec{r},t)=\langle \phi_{\gamma} \,|\hat{E}(\vec{r},t)|\,\phi_{\gamma}\rangle=\\
\sqrt {\frac{\hbar\,\omega_{\vec{k}}}{2\,\varepsilon_0\,V}}\;\;\vec{a}_{\vec{k}}(t)\,\cos[\vec {k}\cdot 
\vec{r}-\omega_{\vec{k}}\,t]\,,
\end{eqnarray}
where
$$
\vec{a}_{\vec{k}}(t)=2\,c_1(t)\,c_2(t)\,\vec{\epsilon}_{\vec{k}}=\frac{1}{\cosh(\frac{2\,t}{\tau})}\;\,
\vec{\epsilon}_{\vec{k}}
$$
describes a bell-shape function of width $\tau$ centered at $t=0$. Within this time span of the absorption process 
$\vec{E}(\vec{r},t)$ oscillates like the electric field of an electromagnetic wave and therefore defines a plane of 
polarization in a completely classical way. It is the vector potential $\vec{A}(\vec{r},t)$ connected with 
$\vec{E}(\vec{r},t)$ through
$$
\vec{E}(\vec{r},t)=-\frac{\partial}{\partial\,t}\,\vec{A}(\vec{r},t)
$$
that goes into the Schr\"odinger equation of the atoms of the optical material and effects the build-up of secondary 
waves whose polarization is hence determined by the polarization of the incoming wave and not by observing it.
\\[0.3cm]
The article by Aspect et al. \cite{Aspect} deals with a situation where pairs of visible photons, correlated in 
their linear polarization, are emitted at a rate of about 5$\times$10$^7 s^{-1}$ from a $^{40}$Ca-source. They are 
monitored by two mirror symmetric instruments facing each other across the source. The setup is schematically shown 
in Fig.1 where the wave trains with which the two photons are associated are indicated together with their plane in 
which $\vec{E}(\vec{r},t)$ oscillates. Just to simplify the ensuing discussions we shall henceforth refer to that 
plane as ``the plane of polarization'' although the latter is conventionally defined as the oscillation plane of the 
magnetic vector.
\begin{figure*}
 \epsfig{file=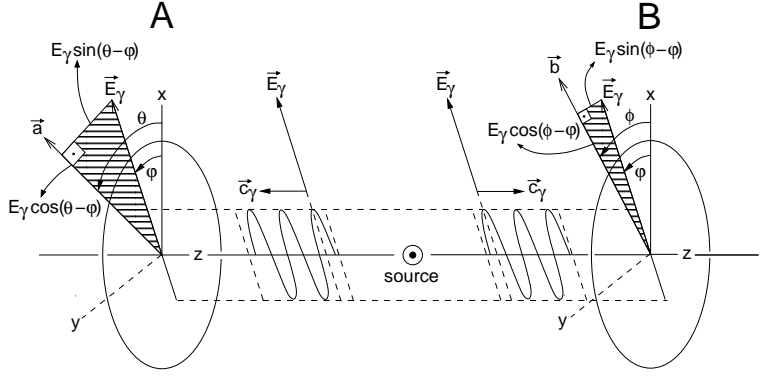, width=10cm}
\caption[EPRB-setup]
{\label{EPRB-setup}Experimental setup for measuring correlated pairs of photons }
\end{figure*}
The amplitude of the electric field vector is denoted by $\vec{E}_{\gamma}$, the vector of light propagation by 
$\vec{c}_{\gamma}$. We define the direction of $\vec{E}_{\gamma}$ to coincide with
$\vec{c}_{\gamma}/c_{\gamma}\times\vec{n}_{\gamma}$ where $\vec{n}_{\gamma}$ denotes the normal of the plane of
polarization. If the latter lies in the $x/y$-plane $\vec{n}_{\gamma}$ points along the $y$-direction.\\
 The vertical planes left and right of the source refer to the entrance faces of the polarizers, stations A and B, 
respectively,  consisting of polarizing cubes that transmit light polarized along a direction $\vec{a}$ on the 
left-hand side and $\vec{b}$ on the right-hand side. Both cubes reflect the perpendicular polarization. Light in the 
direction of transmission and reflection is monitored by photomultipliers and coincidence counter electronics.
\\[0.2cm]
The energy density $u$ within each of the wave trains is given by
\begin{eqnarray}
\label{eqn:energy_density}
u=\varepsilon_0\,E^{2}_{\gamma}\,.
\end{eqnarray}
Hence, in the spirit of the above idea on photons, the quantity
\begin{eqnarray}
\label{eqn:probability_density}
\rho=\frac{u}{\hbar\,\omega_{\gamma}}
\end{eqnarray}
has to be interpreted as the probability density of finding the photon in the respective wave train. \\
If the wave train contains only one photon $\rho$ yields unity on integration over the volume of the wave train. 
Hence, a detector of perfect quantum efficiency in line with the wave train's propagation will definitely fire on 
its arrival if the cross section of the wave train would be the same size or smaller than the sensitive entrance 
face
of the detector. If it has passed through a semi-reflecting/semi-transparent (50-50) beam splitter each of the two 
detectors at the end of the now occurring two beams will fire with only  $50\%$ probability, and there will be no 
coincidences of detector signals. If the wave train contains coherently two photons, the energy density $u$, and 
consequently $\rho$, will be larger by a factor of two so that $\rho_{split}$ in each of the beams behind the 
splitter integrates again to unity. However, this case deserves some more comment:
\\[0.2cm]
As follows from our considerations in connection with Eqs.(\ref{eqn:photonstate}) and (\ref{eqn:oscillating_E}) the 
two-photon state
$$
|\phi_{\gamma}(t)\rangle=c_1(t)\,|n_{\vec{k}}^{(1)}\rangle+c_2(t)\,|n_{\vec{k}}^{(2)}\rangle
$$
where $|n_{\vec{k}}^{(1)}\rangle=2$ can only yield an oscillating electric field $\vec{E}(\vec{r},t)$ driving the 
build-up of the secondary field if $|n_{\vec{k}}^{(2)}\rangle=1$ because the operator $\hat{E}(\vec{r},t)$ contains 
only first-order terms of $\hat{a}_{\vec{k}}$ and $\hat{a}_{\vec{k}}^{\dagger}$. Hence, the two-photon absorption 
$|n_{\vec{k}}^{(1)}\rangle=2\,\rightarrow \,|n_{\vec{k}}^{(2)}\rangle=0$ happens in two consecutive steps giving 
rise to two secondary photons sequentially lined up in real-space. Since each of the two photons has a 50$\%$ chance 
to go to the transmission or reflection channel, there is a 25$\%$ chance that \textbf{both} photons go to the 
transmission or reflection channel. That means, there is a remaining 50$\%$ chance that one photon is transmitted 
and the other one is reflected. Although the probability currents in each of the beams integrates to one photon, 
there will only be a 50$\%$ chance for coincidence signals with the associated detectors.\\
In all what follows Eq.(\ref{eqn:probability_density}) and the implications just discussed will prove to be 
sufficient in analyzing the key experiments on photon correlation.
\\[0.2cm]
As for the experiment by Aspect et al. \cite{Aspect} we only consider two photons that are emitted in a zero-recoil 
mode from an atom and hence travel in opposite directions. The photons are generated in a $^{40}$Ca-cascade 
transition: 4p$^2\rightarrow$ 4p$\,$4s $\rightarrow$ 4s$^2$. The first transition yields a photon of 5513 ${\AA}$ 
wavelength and keeps the dipole axis fixed for the second transition that yields a photon of 4227 ${\AA}$ 
wavelength. Because of the fixed dipole axis the two photons are both linearly polarized with a common plane of 
polarization which is conserved as they propagate. \\
When the left wave train penetrates into the cube (polarizer) of station A it divides up - according to Malus' law - 
into a transmitted wave train with an amplitude $E_{\gamma}\,\cos(\theta-\varphi)$ of the primary electric field and 
into a reflected wave train with amplitude $E_{\gamma}\,\sin(\theta-\varphi)$. Here $\varphi$ denotes the angle that 
the normal of the polarization plane includes with the x-direction of the laboratory-fixed coordinate system whose 
z-axis coincides with the line of propagation of the two photons. The angle $\theta$ is correspondingly the angle 
that the direction $\vec{a}$ of the left polarizer includes with the x-direction. Its counterpart is the angle 
$\phi$ of the polarizer on the right-hand side. Here the respective wave train is decomposed into a transmitted wave 
train with electric field amplitude $E_{\gamma}\,\cos(\phi-\varphi)$  and a reflected wave train the electric field 
amplitude of which is $E_{\gamma}\,\sin(\phi-\varphi)$. \\
To characterize the associated energy densities we use superscripts A and B, respectively, for quantities referring 
to the left- and right-hand side (stations A and B) of the experimental setup. We, furthermore, use subscripts $+$ 
and $-$, respectively, to denote the energy densities of the wave trains transmitted parallel or reflected 
perpendicular to $\vec{a}$ on the left-hand side, and likewise with respect to $\vec{b}$ on the right-hand side.\\
Hence we have according to Eq.(\ref{eqn:energy_density})
\begin{eqnarray*}
 u^{A}_{+}=\varepsilon_0\,E^{2}_{\gamma}\,\cos^2(\theta-\varphi)\,;\quad 
u^{A}_{-}=\varepsilon_0\,E^{2}_{\gamma}\,\sin^2(\theta-\varphi)
\end{eqnarray*}
and correspondingly
\begin{eqnarray*}
 u^{B}_{+}=\varepsilon_0\,E^{2}_{\gamma}\,\cos^2(\phi-\varphi)\,;\quad 
u^{B}_{-}=\varepsilon_0\,E^{2}_{\gamma}\,\sin^2(\phi-\varphi)\,,
\end{eqnarray*}
which yields for the probabilities of finding the respective photons in one of the channels ($+$ or $-$, 
respectively)
\begin{eqnarray}
\label{eqn:ProbA}
 \hat{P}^{A}_{+}(\theta,\varphi)=\frac{u^{A}_{+}}{u^{A}_{+}+u^{A}_{-}}=cos^2(\theta-\varphi)\nonumber \\ 
\hat{P}^{A}_{-}(\theta,\varphi)=\frac{u^{A}_{-}}{u^{A}_{+}+u^{A}_{-}}=\sin^2(\theta-\varphi)
\end{eqnarray}
and
\begin{eqnarray}
\label{eqn:ProbB}
\hat{P}^{B}_{+}(\phi,\varphi)=\frac{u^{B}_{+}}{u^{B}_{+}+u^{B}_{-}}=\cos^2(\phi-\varphi)\nonumber \\ 
\hat{P}^{B}_{-}(\phi,\varphi)=\frac{u^{B}_{-}}{u^{B}_{+}+u^{B}_{-}}=\sin^2(\phi-\varphi)\,.
\end{eqnarray}
The quantities $\hat{P}^{A/B}_{\pm}$ are proportional to the count rates in the associated channels of the 
experimental setup if one generates the photon pairs at a certain rate.\\
If the pairs would all be emitted with $\vec{E}_{\gamma}$ lying in the same plane, but $\varphi$ would be unknown, 
one could determine this angle from the count rates by forming
\begin{eqnarray}
\label{eqn:singleCountRatesA}
\hat{P}^{A}_{+}(\theta,\varphi)-\hat{P}^{A}_{-}(\theta,\varphi)= \qquad \qquad \nonumber \\
\cos^2(\theta-\varphi)-\sin^2(\theta-\varphi)=\cos 2(\theta-\varphi)
\end{eqnarray}
This expression may be viewed as the degree of polarization of the incoming photon or just as its ``polarization'' 
with respect to $\vec{a}$. It equals $+1$ when $\varphi=\theta$ and $-1$ when $\varphi=\theta \pm \pi/2$. In the 
spin-resolved electron scattering at heavy atoms an analogous expression is used to define the polarization of an 
electron beam impinging on a Mott-detector where the difference in the left-right asymmetry of the pertinent 
differential cross section is used in place of $P_{+}-P_{-}$. Incidentally, it is the Mott-detector that is actually 
used in true polarization experiments on massive particles as opposed to the fictional (completely inappropriate) 
Stern-Gerlach magnet commonly discussed in the context of EPRB-experiments. \\
As argued by Einstein in 1949 \cite{Einstein2}, the result A of the measurement on photon 1 should not depend on the 
direction $\vec{b}$ of the polarizer at B acting on photon 2, and B should not depend on $\vec{a}$. Photon 2 is only 
correlated with photon 1 in the sense that its plane of polarization is the same as that of photon 1. That means in 
terms of the count rates at B:
\begin{eqnarray}
\label{eqn:singleCountRatesB}
\hat{P}^{B}_{+}(\phi,\varphi)-\hat{P}^{B}_{-}(\phi,\varphi)=\cos^2(\phi-\varphi)-\sin^2(\phi-\varphi)=\nonumber \\
\cos 2(\phi-\varphi)\quad
\end{eqnarray}
To make sure that each count at A and B refers to the same pair, all four channels are checked by coincidence 
measurements.\\
The differences $\hat{P}^{A/B}_{+}-\hat{P}^{A/B}_{-}$ may be interpreted as probabilities of ``preferential 
detection''
in the ''+''-channel. They attain negative values if the photons are actually detected in the ``-''-channel. Hence,
the expression
$$
\left(\hat{P}^{A}_{+}(\theta,\varphi)-\hat{P}^{A}_{-}(\theta,\varphi) 
\right)\left(\hat{P}^{B}_{+}(\phi,\varphi)-\hat{P}^{B}_{-}
(\phi,\varphi)\right)
$$
may be viewed as the probability of finding the photons at station A preferentially detected in the ``+''-channel 
and
those simultaneously monitored at B with the same preference.
Because of Eqs.(\ref{eqn:singleCountRatesA}) and (\ref{eqn:singleCountRatesB}) this joint probability takes the form
\begin{eqnarray}
\label{eqn:CombinedCountRates}
\left(\hat{P}^{A}_{+}(\theta,\varphi)-\hat{P}^{A}_{-}(\theta,\varphi) 
\right)\left(\hat{P}^{B}_{+}(\phi,\varphi)-\hat{P}^{B}_{-}
(\phi,\varphi)\right)=\nonumber \\
\cos 2\,(\theta-\varphi)\,\cos 2\,(\phi-\varphi)= \nonumber \\
\frac{1}{2}\,\cos 2\,(\theta-\phi)+\frac{1}{2}\,\cos 2\,(\theta+\phi-2\,\varphi)\,.
\end{eqnarray}
Obviously, the first term on the right-hand side becomes $\varphi$-independent only if the polarizations of the 
right and left wave train lie in the same plane.\\
The pairs are emitted such that their planes of polarization are oriented at random. On performing  a
$\varphi$-average of Eq.(\ref{eqn:CombinedCountRates}) over the range $[-\frac{\pi}{2},\frac{\pi}{2}]$ we obtain
\begin{eqnarray}
\label{eqn:AveragedCombinedCountRates}
\overline{\left(\hat{P}^{A}_{+}(\theta,\varphi)-\hat{P}^{A}_{-}(\theta,\varphi) 
\right)\left(\hat{P}^{B}_{+}(\phi,\varphi)-
\hat{P}^{B}_{-}(\phi,\varphi)\right)} \nonumber \\
=\frac{1}{2}\,\cos 2(\theta-\phi).\quad
\end{eqnarray}
We may rewrite the expression on the left-hand side
\begin{eqnarray}
\label{eqn:IndividualCombinedCountRates}
\overline{\left(\hat{P}^{A}_{+}(\theta,\varphi)-\hat{P}^{A}_{-}(\theta,\varphi) 
\right)\left(\hat{P}^{B}_{+}(\phi,\varphi)-\hat{P}^{B}_{-}
(\phi,\varphi)\right)}= \nonumber \\
P_{++}(\theta,\phi)+P_{--}(\theta,\phi)-P_{+-}(\theta,\phi)-P_{-+}(\theta,\phi)\quad \,
\end{eqnarray}
where
\begin{eqnarray}
\label{eqn:Definition}
P_{\pm \pm}(\theta,\phi)=\overline{\hat{P}^{A}_{\pm}(\theta,\varphi)\;\hat{P}^{B}_{\pm}(\phi,\varphi)}\,.
\end{eqnarray}
From Eqs.(\ref{eqn:ProbA}) and (\ref{eqn:ProbB}) we have
\begin{eqnarray}
\label{eqn:one_photon_probability}
P_1=\overline{\hat{P}^{A/B}_{\pm}}=\frac{1}{2}
\end{eqnarray}
which states that the probability of finding, respectively, photon 1 at A or photon 2 at B in one of the two 
channels is equal to $0.5$ on the average. We shall henceforth substitute $\theta$ and $\phi$ by the unit vectors 
$\vec{a}$ and $\vec{b}$ which these angles refer to.\\
The quantities $P_{\pm \pm}(\vec{a},\vec{b})$ represent joint probabilities. Hence, the conditional probability 
$P_{++}^{c}(\vec{a},\vec{b})$ of finding photon 1 at A in the "+" -channel if the companion photon 2 has been 
detected in the "+" -channel at B, is given by
\begin{eqnarray}
\label{eqn:correlatedProbab}
P_{++}^{c}(\vec{a},\vec{b})=\frac{1}{P_{1}}\,P_{++}(\vec{a},\vec{b})\,.
\end{eqnarray}
The quantities $P_{\pm \pm}^{c}(\vec{a},\vec{b})$ for other sign combinations are defined analogously. Hence, if we 
set
\begin{eqnarray}
\label{eqn:DefCorrelationFactor}
E=\frac{1}{P_1}\,\overline{\left(\hat{P}^{A}_{+}(\theta,\varphi)-\hat{P}^{A}_{-}(\theta,\varphi) 
\right)\left(\hat{P}^{B}_{+}(\phi,\varphi)-
\hat{P}^{B}_{-}(\phi,\varphi)\right)} \nonumber \\
=P_{++}^{c}(\vec{a},\vec{b})+P_{--}^{c}(\vec{a},\vec{b})-P_{+-}^{c}(\vec{a},\vec{b})-P_{-+}^{c}(\vec{a},\vec{b})
\quad \quad
\end{eqnarray}
and use the above equations from (\ref{eqn:AveragedCombinedCountRates}) to (\ref{eqn:correlatedProbab}), we obtain
\begin{eqnarray}
\label{eqn:DefCorrelationFactor}
E=\cos2(\theta-\phi)\equiv\cos 2(\vec{a},\vec{b})\,.
\end{eqnarray}
The quantity $E$ constitutes the so-called correlation coefficient of the measurement on the two photons, and 
exactly this equation is fully confirmed by the experiment. By hindsight this may also be seen as justifying our 
assumption on the uniform distribution of the angle $\varphi$.\\
As indicated in Fig.1 this angle describes the orientation of the plane of polarization in which the electric field 
of the wave trains oscillates. It hence represents, in the spirit of the EPR-article, an element of physical reality 
that allows one to predict with certainty how the wave trains impinging on their respective polarizers divide up 
into secondary wave trains with energy densities $u_{+}$ and $u_{-}$. Admittedly, these densities correlate only 
with the probabilities of transmission and reflection for an incoming single photon and thus represent typical 
elements of uncertainty. However, the actual choice made by the individual photon cannot be predicted by any of the 
existing theories.
\\
It should also clearly be stated that quantum theory cannot make any definite prediction on the position of a photon
within its associated wave train. The position along its line of propagation is only determined up to the length of 
the wave train. On the other hand, it appears to be out of the question that the distance of the two photons from 
the source remains exactly equal along this line of propagation and therefore represents another element of physical 
reality not described by quantum theory.\\
Furthermore, it is obvious from our treatment that a photon traversing the polarizer at A possesses now a 
polarization $\vec{E}_{\gamma}\parallel \vec{a}$ that differs from that of the original (incoming) photon if 
$\theta-\varphi \not=0$. From the viewpoint of quantum theory, which is free from hidden parameters, a photon can  
have a definite polarization only after it has left the polarizer, which represents a rather implausible credo 
because the occurrence of this ``definite polarization'' can only be explained by assuming a classical functioning 
of the polarizer.
\\[0.2cm]
Quantum entanglement constitutes a particularly puzzling feature of describing two- or multi-photon correlation.  
This becomes even more apparent in the case of pairs that consist of identical massive particles with spin in an 
entangled singlet state. The latter implies  equal probabilities of finding either particle with either spin 
orientation at station A and B, and only after the detector at A has measured a particle with ``spin-up'', the 
detector at B yields definitely a coincidence signal that correlates with  ``spin-down'' and vice versa. This 
amounts to an instantaneous \textit{non-local} intercommunication between even very distant stations A and B. Hence, 
if one denies the reality of single events as described by our approach and adheres to the idea of quantum 
mechanical completeness defined by a state vector of the system, one is forced to ascribe the measurement a decisive 
influence on the system under study and to put up with \textit{spooky action at a distance}.\\
As one follows the various steps in deriving Eq.(\ref{eqn:AveragedCombinedCountRates}) one recognizes that it 
describes basically a classical behavior of electromagnetic waves. This can be seen by considering a radio wave 
source that consists of a Hertzian oscillator of frequency $\omega_{\gamma}$ whose dipole axis is intermittently 
turned at random  such that the angle $\varphi$ it includes with the $x$-axis becomes uniformly distributed over a 
unit circle in the $x/y$-plane. We assume that there are polarizers positioned at A and B as depicted in Fig.1. They 
may consist of a frame similar to a tennis racket that contains only one set of parallel (superconducting) strings 
with the second orthogonal set of strings missing. The plane of these strings is tilted by 45$^{0}$ against the 
$z$-axis. The axis of these polarizer rackets includes angles $\theta$ and $\phi$, respectively, with the $x$-axis. 
A radio wave whose electric field vector $\vec{E}_{\gamma}$ is parallel to the strings will be reflected by 
90$^{0}$, and it is transmitted if $\vec{E}_{\gamma}$ is perpendicular to the strings. If $\vec{E}_{\gamma}$ 
includes an angle $\varphi$ with the $x$-axis, the relative intensities $I^{A/B}_{\pm}$ for transmission and 
reflection are given by Eqs.(\ref{eqn:ProbA}) and (\ref{eqn:ProbB}).
Hence, if one averages these intensities over a sufficiently long time and forms
$I_{\pm \pm}=I^{A}_{\pm}\,I^{B}_{\pm}$ one obtains in complete analogy to Eq.(\ref{eqn:DefCorrelationFactor})
\begin{eqnarray*}
E=2\,[I_{++}+I_{--}-I_{+-}-I_{-+}]=\cos2(\theta-\phi)\,.
\end{eqnarray*}
\begin{eqnarray*}
\qquad \qquad \qquad \qquad
\end{eqnarray*}
\section{ Bell's disputable message on the EPRB-experiment}\label{Bell-message}
Since the experiment by Aspect et al. \cite{Aspect} refers explicitly to the EPRB-experiment and to John Bell's well 
publicized analysis \cite{Bell1} of it, we prefer to discuss the essence of Bell's considerations from an article 
that appeared 17 years later \cite{Bell2} and follows closely our line of thought. Bell considers pairs of neutral 
particles with opposite spins that run in opposite directions through Stern-Gerlach magnets each placed in the 
direction of flight of the respective particle. He allows for an angle $\varphi$ that the particle-spin includes 
with a laboratory-fixed x-axis, similar to the angle we introduced in the previous section. This analogy is carried 
further in that he lets the field gradient of the Stern-Gerlach magnets include angles $\theta$ and $\phi$ , 
respectively, with that x-axis. These angles may be replaced by unit vectors $\vec{a}$ and $\vec{b}$ whose 
directions are defined by the angles. Completely in accord with our line of thought he considers the angle $\varphi$ 
to be a random variable associated with a very large set of pairs ejected one after the other from the same source. 
\\
The critical and eventually disastrous point of departure is marked by his assumption that the particle at station A 
is monitored with \textit{certainty} in the ``up '' (or ``+'')-channel and remains constant at this probability $1$ 
as long as $-\frac{\pi}{2}<\theta-\varphi<\frac{\pi}{2}$, and it goes definitely to the ``down'' (or ``-'')-channel 
if $\frac{\pi}{2}<\theta-\varphi<\frac{3\,\pi}{2}$. The same is assumed to apply to the particle monitored at B. If 
he now runs through a sufficiently large set of pairs and averages over $\varphi$, he obtains for the probability 
for ``up, up''- or ``down, down''-events
\begin{eqnarray}
\label{eqn:Bell_up_up}
P^{f}_{++}=P^{f}_{--}=\frac{|\theta-\phi|}{2\,\pi}\,,
\end{eqnarray}
and for ``up, down''-events
\begin{eqnarray}
\label{eqn:Bell_down_down}
P^{f}_{+-}=P^{f}_{-+}=\frac{1}{2}-\frac{|\theta-\phi|}{2\,\pi}\,.
\end{eqnarray}
Here we have introduced a superscript ``f'' ( for ``fermion'') to indicate the reference to massive particles with 
half-integer spin.\\
The two equations yield a linear dependence of $E$ on $|\theta-\phi|$, viz.
\begin{eqnarray}
\label{eqn:E_Bell1}
E^{Bell}=P^{f}_{++}+P^{f}_{--}-P^{f}_{+-}-P^{f}_{-+}= \nonumber \\
 -1+2\,\frac{|\theta-\phi|}{\pi}\,.
\end{eqnarray}
If one translates Bell's considerations into the case of photons and applies them in particular to the experiment by 
Aspect et al. \cite{Aspect}, the corresponding linear dependence reads
\begin{eqnarray}
\label{eqn:E_Bell2}
E^{Bell}=1-2\,\frac{2\,|\theta-\phi|}{\pi}
\end{eqnarray}
which differs fundamentally from our result (\ref{eqn:DefCorrelationFactor})
$$
E=\cos 2(\theta-\phi)
$$
although the two expressions agree for three particular values of $|\theta-\phi|$, viz. $0^{\circ}, 45^{\circ}$ and 
$90^{\circ}$.
\\[0.2cm]
As for the photon case his assumption is clearly inadmissible: a photon whose associated field vector 
$\vec{E}_{\gamma}$ makes an angle $\theta-\varphi$ with the polarizer at station A, has already a non-vanishing 
probability for going to the ``-''-channel as soon as $\theta-\varphi$ differs from zero. This becomes obvious from 
tracing the origin of our results back to Eqs.(\ref{eqn:ProbA}) and (\ref{eqn:ProbB}).\\
Interestingly, Clauser and Horne \cite{Clauser} discuss a model for the correlation of photon counts that bears some 
resemblance to our approach, but their ad hoc-assumption on the rates $P^{A}_{+}$ and $P^{B}_{+}$ such that 
$P^{A}_{+}-P^{B}_{+}$ agrees with the experiments lacks any physical foundation.
\\[0.2cm]
Bell's highly recognized inequality (against which Aspect et al. checked their experiment) rests on the above 
assumption that leads to Eq.(\ref{eqn:E_Bell2}). Since the considered mechanism of sorting particles into two groups 
``+'' and ``-'' is definitely unrealistic for photons, a test of hidden parameter theories that rest on Bell's 
inequality is meaningless. In discussing the EPRB-experiment and summarizing the message of the above equations 
(\ref{eqn:Bell_up_up}) and (\ref{eqn:Bell_down_down}) in his inequality, Bell wanted to demonstrate that one 
\textit{cannot} explain the ``exact quantum mechanical result'' \textit{without} assuming action at a distance. A 
purely \textit{local} mechanism sorting the members of particle pairs into ``+'' and ``-'' channels would definitely 
lead to a confirmation of his inequality. This is obviously not true for correlated photon pairs as we have shown in 
Section \ref{correlated_pairs} where we derive the correct  photon correlation factor (which violates Bell's 
inequality) assuming a purely local mechanism of particle separation. One might argue that this only reflects the 
inadequacy of Bell's line of reasoning for correlated photon pairs. However, as we shall demonstrate in the ensuing 
section, one can just as well explain the ``exact quantum mechanical result'' for the true EPRB-experiment by 
assuming again a purely local mechanism.
\section{Locality vs. Non-Locality}\label{Locality}
 We start with the quantum mechanical expression for what is believed to be connected to the count rate in the four 
channels of an EPRB-setup where the counting is organized in complete analogy to the experiment by Aspect et al. 
\cite{Aspect}:
\begin{eqnarray}
\label{eqn:EPRB_countrate}
E=\langle \Psi_0|\vec{\sigma}_{A}\cdot \vec{a} \otimes \vec{\sigma}_{B}\cdot 
\vec{b}|\Psi_0\rangle=-\cos(\vec{a},\vec{b})
\end{eqnarray}
which describes the statistical correlation of monitoring one of the two particles at station A with a spin 
component in line with  $\vec{a}$ and the counterpart of the two particles with a spin component in line with 
$\vec{b}$.
Here $\Psi_0$ stands for an entangled state (``Bell state'')
\begin{eqnarray}
\label{eqn:EntangledState}
\Psi_0=\frac{1}{\sqrt{2}}\,[S_A(\vec{r}_1)\,e^{-i\vec{k}\cdot \vec{r}_1}\,\left( {1 \atop 0}\right)\otimes 
S_B(\vec{r}_2)\,e^{i\vec{k}\cdot \vec{r}_2}\,\left( {0 \atop 1} \right)-\;  \nonumber \\
S_A(\vec{r}_2)\,e^{-i\vec{k}\cdot \vec{r}_2}\,\left( {0 \atop 1}\right)\otimes S_B(\vec{r}_1)\,e^{i\vec{k}\cdot 
\vec{r}_1}\,\left( {1 \atop 0}\right)]\,. \quad
\end{eqnarray}

To retain the familiar notation for the Pauli spin matrices $\vec{\sigma}_A, \vec{\sigma}_B$, we have rotated the 
coordinate system such that $z\rightarrow x, -y\rightarrow y$ and $x\rightarrow z$. Hence $\hbar\,\vec{k}$ denotes 
the particle momentum in the x-direction, and $S_{A}(\vec{r}); S_{B}(\vec{r})$ represent smooth, real-valued 
functions which vanish around the source and reach well into the Stern-Gerlach magnet at station A and B, 
respectively. They are sizeably different from zero only in a narrow cylinder around the x-axis and are essentially 
constant within this domain. Their squares integrate to unity. The unit vectors $\vec{a}$ and $\vec{b}$ are given by
\begin{eqnarray*}
\vec{a}=(0, -\sin \theta,\cos \theta)\,; \quad \vec{b}=(0, -\sin \phi, \cos \phi)\,.
\end{eqnarray*}
Expression (\ref{eqn:EPRB_countrate}) follows simply from taking the expectation value of
\begin{eqnarray*}
\vec{\sigma}_{A}\cdot \vec{a}\otimes \vec{\sigma}_{B}\cdot \vec{b}=\sigma_y \otimes \sigma_y\,\sin \theta\,\sin \phi 
+\sigma_z \otimes \sigma_z\,\cos\theta\,\cos\phi \;\, \\ \nonumber
-\sigma_y \otimes \sigma_z\,\sin \theta\,\cos \phi -\sigma_z \otimes \sigma_y\,\cos\theta\,\sin\phi\,.
\end{eqnarray*}
In the spirit of our derivation that led to the set of Eqs.(\ref{eqn:AveragedCombinedCountRates}) to 
(\ref{eqn:correlatedProbab}) $E$ can alternatively be cast as
\begin{eqnarray}
\label{eqn:Mycountrate}
E=\frac{1}{P_1}\,\overline{\left(\hat{P}^{A}_{+}(\theta,\varphi)-\hat{P}^{A}_{-}(\theta,\varphi) 
\right)\left(\hat{P}^{B}_{+}(\phi,\varphi)-\hat{P}^{B}_{-}
(\phi,\varphi)\right)}\nonumber \\
=P_{++}^{c}(\vec{a},\vec{b})+P_{--}^{c}(\vec{a},\vec{b})-P_{+-}^{c}(\vec{a},\vec{b})-P_{-+}^{c}(\vec{a},\vec{b})
\quad\;
\end{eqnarray}
where $P_{\pm \pm}^{c}=2\,P_{\pm \pm}$ and $P_{\pm \pm}$ is defined as before
\begin{eqnarray}
\label{eqn:Definition}
P_{\pm \pm}(\vec{a},\vec{b})=\overline{\hat{P}^{A}_{\pm}(\vec{a},\vec{b})\,\hat{P}^{B}_{\pm}(\vec{a},\vec{b})}
\end{eqnarray}
We are now in the position to discuss E as in the previous case of correlated photons. That means, we do not discuss 
pairs in entangled states but rather individual pairs, more precisely: a set of subsequently emitted pairs of 
particles, in which one of the particles is definitely moving toward A, the other one toward B.  Correspondingly, we 
start again by considering the probability $\hat{P}^{A}_{+}(\theta,\varphi)$ of finding the particle at A with spin 
up if it has entered the station in a state
\begin{eqnarray}
\psi^{A}(\vec{r})=S_{A}(\vec{r})\,e^{-i\,\vec{k}\cdot \vec{r}}\,\left[\,\cos\frac{\varphi}{2}\,\left( {1 \atop 
0}\right)-i\,\sin\frac{\varphi}{2}\,\left( {0 \atop 1}\right)\right]
\end{eqnarray}
whose spin encloses an angle $\varphi$ with the z-axis. When it has been monitored in the ``up''-channel its state 
is  given by
\begin{eqnarray}
\psi^{A}_{+}(\vec{r})=S_{A}(\vec{r})\,e^{-i\,\vec{k}\cdot \vec{r}}\,\left[\,\cos\frac{\theta}{2}\,\left( {1 \atop 
0}\right)-i\,\sin\frac{\theta}{2}\,\left( {0 \atop 1}\right)\right]\,.
\end{eqnarray}
From this we may determine the transition probability
\begin{eqnarray*}
|\langle \psi^{A}_{+}(\vec{r})|\psi^{A}(\vec{r})\rangle|^2=\left|\cos \frac{\varphi}{2}\,
\cos \frac{\theta}{2}+
\sin \frac{\varphi}{2}\,\sin \frac{\theta}{2}\right|^2\,,
\end{eqnarray*}
that is
\begin{eqnarray}
\label{eqn:Prob_up}
\hat{P}^{A}_{+}(\theta,\varphi)=\cos^{2}(\frac{\theta-\varphi}{2})\,.
\end{eqnarray}
Likewise one obtains
\begin{eqnarray*}
\hat{P}^{A}_{-}(\theta,\varphi)= \left|-i\,\cos\frac{\varphi}{2}\,
\sin \frac{\theta}{2}+i\,
\sin \frac{\varphi}{2}\,\cos \frac{\theta}{2}\right|^2=\\
\sin^{2}(\frac{\theta-\varphi}{2})\,.
\end{eqnarray*}
Hence
\begin{eqnarray*}
\hat{P}^{A}_{+}(\theta,\varphi)-\hat{P}^{A}_{-}(\theta,\varphi)=\cos^{2}(\frac{\theta-\varphi}{2})-\sin^{2}
(\frac{\theta-\varphi}{2})=\\
\cos(\theta-\varphi)\,.\quad
\end{eqnarray*}
The corresponding expression for station B reads
\begin{eqnarray*}
\hat{P}^{B}_{+}(\theta,\varphi)-\hat{P}^{B}_{-}(\phi,\varphi)= \qquad \qquad \qquad \qquad  \\
 \left[\sin^{2}(\frac{\phi-\varphi}{2})-\cos^{2}(\frac{\phi-\varphi}{2})\right]=-\cos(\phi-\varphi)\,\qquad
\end{eqnarray*}
If we insert this into Eq.(\ref{eqn:Mycountrate}) we obtain
\begin{eqnarray}
\label{eqn:SpinCorrelation}
E=2\,\overline{\left(\hat{P}^{A}_{+}(\theta,\varphi)-\hat{P}^{A}_{-}(\theta,\varphi) 
\right)\left(\hat{P}^{B}_{+}(\phi,\varphi)-\hat{P}^{B}_{-}
(\phi,\varphi)\right)}\nonumber\\
=-\cos(\theta-\phi)\qquad
\end{eqnarray}
which may alternatively be written
\begin{eqnarray}
\label{eqn:SpinCorrelation_2}
E=-\vec{a}\cdot \vec{b}
\end{eqnarray}
To make contact to Bell's notation in his seminal paper \cite{Bell1} we identify the angle $\varphi$ in 
Eq.(\ref{eqn:SpinCorrelation}) with his single parameter $\lambda$. Furthermore, we have to relate his functions 
$A(\vec{a},\lambda)$ and $B(\vec{b},\lambda)$ to our expressions $\hat{P}^{A/B}_{\pm}$
$$
A(\theta,\varphi)=A(\vec{a},\lambda)=
\sqrt 2\,[\hat{P}^{A}_{+}(\vec{a},\lambda)-\hat{P}^{A}_{-}(\vec{a},\lambda)]
$$
and
$$
B(\phi,\varphi)=B(\vec{b},\lambda)=
\sqrt 2\,[\hat{P}^{B}_{+}(\vec{b},\lambda)-\hat{P}^{B}_{-}(\vec{b},\lambda)]\,.
$$
The two spins and the line along which the two particles propagate span a plane whose normal encloses the angle 
$\varphi$ with the $z$-axis. We assume that $\varphi$ is uniformly distributed over the unit circle if the emission 
of pairs is repeated sufficiently often. Hence, we have for Bell's probability distribution
$$
\rho(\lambda)=\frac{1}{\pi}\,.
$$
Eq.(\ref{eqn:SpinCorrelation}) then takes the form
$$
E\equiv 
P^{Bell}(\vec{a},\vec{b})=\int_{-\pi/2}^{+\pi/2}\rho(\lambda)\,A(\vec{a},\lambda)\,B(\vec{b},\lambda)\,d\lambda 
=-\vec{a}\cdot \vec{b}\,,
$$
where we have inserted our result (\ref{eqn:SpinCorrelation_2}) on the right-hand side. It is exactly this equality 
that is fundamentally questioned in Bell's article: \textit{``It will be shown that this is not possible.''}\\
He considers the property of $A(\vec{a},\lambda)$ to be independent of the setting $\vec{b}$ at station B and 
$B(\vec{b},\lambda)$ to be independent of $\vec{a}$ as defining the hypothesis of \textit{locality} and restates the 
above assertion in a different context \cite{Bell3}: \textit{``With these local forms, it is {\bf not} possible to 
find functions $A$ and $B$ and a probability distribution $\rho$ which give the correlation $-\vec{a}\cdot 
\vec{b}$.''}\\
As we have already stated in Section \ref{Bell-message}, he is led to this contradictory conclusion by his 
completely unfounded assumption on the $\varphi$-dependence of monitoring the particles at stations A and B.
\\[0.2cm]
To leave no shade of uncertainty, we emphasize again that our result Eq.(\ref{eqn:SpinCorrelation_2}) is
based
on the idea of what is commonly termed ``local realism''. It is therefore exceedingly puzzling that even the most
recent articles on this subject choose the Stern-Gerlach twin setup to illustrate the idea of ``local realism'' by
explaining:\\
\textit{``Yet, for the two systems oriented in parallel (i.e. $\theta=\phi=0$), once the, say, ``up'' detector of
particle
1 has fired, we know \textbf{with certainty} that the ``down'' detector of particle 2 will register on the other 
side and vice
versa.''}(Zeilinger \cite{Zeilinger(0)})
\\
At some other place of the article one reads:
\\[0.2cm]
\textit{''....neither }[particle] \textit{has a well defined spin before it is measured.''}
\\[0.2cm]
As we have stated at the outset, the latter assertion is without foundation and actually refuted by our derivation 
whose
consistency with the experiment rests on the contrary assumption.
As regards the first statement, it has to be recalled that the probability of detecting particle 1 in the 
``up''-channel is according to Eq.(\ref{eqn:Prob_up})
$\hat{P}^{A}_{+}=\cos^2(\frac{1}{2}\,\varphi)$, and we have also for particle 2 in the ``down''-channel
$\hat{P}^{B}_{-}=\cos^2(\frac{1}{2}\,\varphi)$.
Clearly, if $\varphi\not= 0$ and therefore $\cos^2(\frac{1}{2}\,\varphi)<1$ particle 1 has a non-vanishing 
probability
(yet smaller than unity) to be registered in the ``up''-channel. But since we also have 
$P^{B}_{-}=\cos^2(\frac{1}{2}\,\varphi)<1$,
and consequently $P^{B}_{+}=\sin^2(\frac{1}{2}\,\varphi)>0$ there is no guarantee that particle 2 will be detected 
in the ``down''-channel.\\
The analogous experiment with photons is similarly commented in the literature and the conclusions are similarly
besides the point.
\section{Quantum teleportation}\label{Teleportation}
Experiments that are believed to demonstrate ``quantum teleportation'' have during the past decade gained 
considerable attention. We mention here only the ground breaking-work by Zeilinger and associates \cite{Zeilinger} 
and refer the reader to the article by Greenberger et al. \cite{Greenberger} for a more detailed exposition of the 
world view of this school of thought. The experimental setup is schematically shown in Fig.2 where a birefringent 
non-linear $\beta$-barium borate (BBO)-crystal acts as a parametric down-conversion source of photons pairs, each 
consisting of a so-called ``signal'' and an ``idler''-photon. They are associated with the classical ordinary and 
extraordinary beam (``o-beam'' and ``e-beam''), the electric field vector of which oscillates, respectively, 
perpendicular to the principal plane or within it. That plane is spanned by the optic axis of the uniaxial crystal 
and the direction of the incoming UV-pulse (of 200$\,fs$ length and 390$\,nm$ wavelength) and is perpendicular to 
the drawing plane.  The angle between these two directions is about 50$^{\circ}$. Because it differs from zero and 
90$^{\circ}$, the directions of all signal photons lie on a cone with an axis in the principal plane. The directions 
of the idler photons span another cone whose axis lies also in the principal plane. The two cones intersect along 
two lines that enclose an angle of about 6$^{\circ}$. They span a plane which coincides with the drawing plane in 
Fig.2.  Photons of pairs that are emitted along these lines, labeled (2) and (3), are in a particular way 
``entangled''. That means in our interpretation: their polarization is no longer uniquely determined by either lying 
in the principal plane or perpendicular to it. The electric field vector $\vec{E}_{\gamma_{2}}$ of photon 2 may now 
enclose any angle with the principal plane. But if one decomposes $\vec{E}_{\gamma_{2}}$ into components 
$\vec{E}_{\gamma_{2}}^{\parallel};\,\vec{E}_{\gamma_{2}}^{\perp}$ parallel and perpendicular to the principal plane 
and performs the same decomposition on $\vec{E}_{\gamma_{3}}$ of photon 3, one finds 
$\vec{E}_{\gamma_{2}}^{\parallel}=\vec{E}_{\gamma_{3}}^{\perp}$ and 
$\vec{E}_{\gamma_{2}}^{\perp}=\vec{E}_{\gamma_{3}}^{\parallel}$. That means: if the normal of the polarization plane 
of photon 2 makes an angle $\varphi$ with the normal of the drawing plane, the normal of the corresponding plane of 
photon 3 makes an angle $\varphi+\frac{\pi}{2}$ with the normal of the drawing plane. (The latter is, incidentally, 
also common to all other beams shown in Fig.2.) Hence, the polarizations of these two photons are always orthogonal.
\begin{figure*}
\epsfig{file=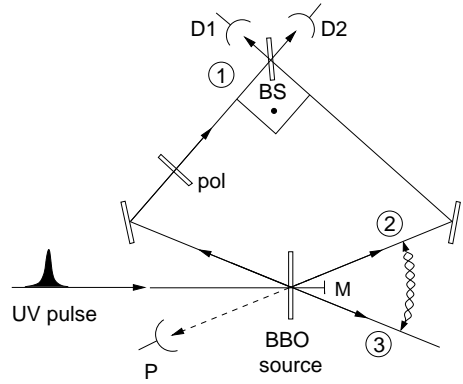,width=6cm}
\caption[photon_entanglement]
{\label{photon_entanglement}Experimental setup for ``quantum teleportation''}
\end{figure*}
As in the previous sections, the angle $\varphi$ represents a random variable that attains a certain value at each 
emission act. \\
The BBO-crystal does not completely absorb the incoming UV-pulse. The residual pulse leaving the crystal impinges on 
a small mirror (M) in front, is reflected and thereby forced to traverse the crystal another time in the opposite 
direction. On its way it generates another pair of ``entangled'' photons. The photon impinging on the detector ``p'' 
is used to ensure the occurrence of its companion photon in the direction (1). That photon is reflected into the 
polarizer ``pol'' where it is converted into a photon whose electric field vector points in some set direction 
$\vec{a}$. At the beam splitter ``BS'' the associated wave train joins at an angle of $\pi/2$ with that of the 
photon originally propagating along (2). The non-polarizing 50-50 beam splitter consists of a half-silvered mirror. 
We denote the electric field amplitudes of wave train (1) and (2) in front of the mirror by $\vec{E}_{\gamma_{1}}$ 
and $\vec{E}_{\gamma_{2}}$, respectively. The resulting amplitude $\vec{E}_{\gamma}^{(D1)}$ of the joint wave train 
traveling toward detector D1 is given by
$$
\vec{E}_{\gamma}^{(D1)}=\vec{E}_{\gamma_{1}}^{(r)}+\frac{1}{\sqrt{2}}\,\vec{E}_{\gamma_{2}} \quad \mbox{where} \quad 
|\vec{E}_{\gamma_{1}}^{(r)}|= \frac{1}{\sqrt{2}}\,|\vec{E}_{\gamma_{1}}|\,\cos\phi
$$
with $\phi$ denoting a possible phase difference between the two wave trains and $\vec{E}_{\gamma_{1}}^{(r)}$ refers 
to the electric field amplitude of the reflected wave train. \\
Correspondingly, we have at the second detector D2
$$
\vec{E}_{\gamma}^{(D2)}=\vec{E}_{\gamma_{2}}^{(r)}+\frac{1}{\sqrt{2}}\,\vec{E}_{\gamma_{1}} \quad \mbox{where} \quad 
|\vec{E}_{\gamma_{2}}^{(r)}|= \frac{1}{\sqrt{2}}\,|\vec{E}_{\gamma_{2}}|\,\cos\phi\,.
$$
After time averaging the associated energy densities are given by
$$
u^{(D1)}= \frac{1}{2}\,u_1 +\frac{1}{2}\,u_2 
+\varepsilon_0\,|\vec{E}_{\gamma_{1}}|\,|\vec{E}_{\gamma_{2}}|\,\cos\phi\,\cos(\alpha+\pi)
$$
where $\alpha$ is the angle between $\vec{E}_{\gamma_{1}}$ and $\vec{E}_{\gamma_{2}}$. There is a phase jump of 
$\pi$ as the wave train connected with $\vec{E}_{\gamma_{1}}$ continues its propagation after metallic reflection.\\
Likewise we obtain
$$
u^{(D2)}= \frac{1}{2}\,u_1 +\frac{1}{2}\,u_2 
-\varepsilon_0\,|\vec{E}_{\gamma_{1}}|\,|\vec{E}_{\gamma_{2}}|\,\cos\phi\,\cos(\alpha+\pi)\,.
$$
Thus, on integrating the densities $u^{(D1/D2)}$ over the volumes $V_{wtr}^{(D1)}\,; V_{wtr}^{(D2)}$ of the wave 
trains propagating toward D1 and D2, respectively, we get
\begin{eqnarray}
\label{eqn:Beamsplitterenergies}
E^{(D1/D2)}=\frac{1}{2}\,E_{1}+\frac{1}{2}\,E_{2}\pm \Delta E
\end{eqnarray}
where
$$
E_{1/2}=u_{1/2}\,V_{wtr}^{(D1/D2)}
$$
and
$$
 \Delta E= \varepsilon_0\,|\vec{E}_{\gamma_{1}}|\,|\vec{E}_{\gamma_{2}}|\,\cos\phi\,\cos(\alpha+\pi)\,
V_{wtr}^{(D1)}\,.
$$
Eq.(\ref{eqn:Beamsplitterenergies}) ensures the conservation of energy:
$$
E^{(D1)}+E^{(D2)}=E_1 +E_2\,.
$$
Hence, if the two photons possess orthogonal polarizations, that is when
$$
\alpha=\frac{\pi}{2}\,,
$$
the probability $E^{(D1)}/\hbar\,\omega_{\gamma}$ of finding one of the two photons at D1 becomes equal to the 
probability $E^{(D2)}/\hbar\,\omega_{\gamma}$ at D2. That means: if there is a coincidence of the signals from the 
two detectors, the two photons must have had orthogonal polarizations. Since the polarization of photon 2 is 
orthogonal to that of photon 3, the latter must have the polarization of photon 1 after the polarizer.\\
All that is demonstrated by this experiment is that the D1/D2-coincidence electronics picks out of the 
$\varphi$-dependent set of pairs that consist of photons 2 and 3 just a photon 3 whose polarization is parallel to 
that of photon 1 after the polarizer. From our point of view there is nothing that would indicate a magic ``quantum 
teleportation'' of polarization from photon 1 to photon 3.
\\[0.2cm]
Since teleportation has gained considerable popularity in the recent past we want to illustrate the cogency of our 
conclusion by simplifying our line of argument in a thought-experiment using essentially the same setup:\\
We replace each pair of mutually orthogonal polarized photons, propagating along the beams (2) and (3), by a pair of 
``color-correlated'' photons, which means, they are associated with two different frequencies that belong to two 
complementary colors adding up to white. The frequencies change statistically from pair to pair so as to cover the 
full range of the visible spectrum. We substitute the polarizer ``pol'' by a filter that is set at some wavelength 
which may correspond to ``yellow'', for example. The beam splitter is replaced by a detector that fires at white 
balance, that is, when the color of the photon arriving along path (2) is the white supplement of the yellow 
``passenger photon'' (1). If this is the case, the photon travelling along path (3) must be yellow as well. Clearly 
there is no teleportation of color from photon (1) to (3), because the color of the latter is already set before 
photon (1) reaches the detector.
\section{Multi-photon entanglement}\label{Multi-photon}
Spontaneous parametric down-conversion has so far proved to be the most effective source for polarization entangled 
photon pairs. The  entanglement of more than two photons has been shown to be feasible by exposing a BBO-crystal to 
short pulses of ultraviolet light as in the experiments on ``quantum teleportation'' discussed in the previous 
section. Multi-photon entanglement is believed to yield the ultimate criterion for distinguishing local hidden 
variable theories from quantum mechanics. The purpose of this section is to again cast  doubt on the validity of 
this belief. The setup of recent experiments dealing with the entanglement of 4 photons is as follows (s. for 
example the article by Weinfurter and associates \cite {Weinfurter}):
\\[0.2cm]
Similar to the 2 forward-photons considered in the preceding section, the 4 photons are emitted into modes $a$ and 
$b$ defining two forward beams which coincide again with the two lines of intersection of the cones for signal and 
idler photons. Each beam is split up into two orthogonal beams by a non-polarizing beam splitter. Hence, after the 
two beam splitters one has 4 beams, all lying in the plane spanned by the original beams $a$ and $b$ which, together 
with the incident UV beam, form a y-shaped array.  Each of the 4 beams enters into a polarizing beam splitter of the 
kind that was used by Aspect et al. \cite{Aspect}. Photons that are transmitted or reflected by these polarizing 
cubes are monitored by eight photon counters all of which are interconnected and checked by means of an 
eight-channel multi-coincidence logic. Each of the four cubes can be rotated around the incoming (and transmitted) 
beam. The respective rotation angles $\phi_{\,a}, \phi_{\,a'}, \phi_{\,b}, \phi_{\,b'}$ are at reference zero when 
the reflected beams lie in the $a/b$-plane.\\
If one applies a reasoning similar to that of the previous section, ``entanglement'' of the 4 photons means:\\
The two photons of  the ``$a$''-beam have the same plane of polarization in common before they impinge on the 
non-polarizing beam splitter , and this applies also to the two photons of the ``$b$''-beam. If the normal 
directions of these two planes of polarization enclose an angle of $90^{\circ}$, all four photons are correlated 
(``entangled''). By contrast, if this angle turns out to be a random variable, one is dealing with two uncorrelated 
pairs of correlated (``entangled'') photons.
\\[0.2cm]
Following the same line of reasoning as in Section \ref{correlated_pairs} we obtain for the corresponding 
probabilities ($\propto$ count rates)
\begin{eqnarray}
\label{eqn:FourPhoton}
\hat{P}^{\,a}_{+}=\cos^2(\phi_{a}-\varphi_{a})\,;\quad \hat{P}^{\,a}_{-}=\sin^2(\phi_{a}-\varphi_{a})\quad \\
\hat{P}^{\,a'}_{+}=\cos^2(\phi_{a'}-\varphi_{a})\,;\quad \hat{P}^{\,a'}_{-}=\sin^2(\phi_{a'}-\varphi_{a})\,,
\end{eqnarray}
and analogous expressions for the $b$-beam where the expressions with $\cos^2$ and $\sin^2$ are just interchanged.
If we were only dealing with the two photons of the ``$a$''-beam, the associated correlation factor would be given 
by
\begin{eqnarray}
\label{eqn:CorrelFactorFourPhoton_a}
E_{a}(\varphi_a)=2\,(\hat{P}^{\,a}_{+}-\hat{P}^{\,a}_{-})\,(\hat{P}^{\,a'}_{+}-\hat{P}^{\,a'}_{-})=\quad \\ 
\nonumber
\cos 2(\phi_{a}-\phi_{a'})+\cos 2(\phi_{a}+\phi_{a'}-2\,\varphi_a)\,.
\end{eqnarray}
and likewise for the ``b''-beam
\begin{eqnarray}
\label{eqn:CorrelFactorFourPhoton_b}
E_{b}(\varphi_b)=2\,(\hat{P}^{\,b}_{+}-\hat{P}^{\,b}_{-})\,(\hat{P}^{\,b'}_{+}-\hat{P}^{\,b'}_{-})=\quad \\ 
\nonumber
\cos 2(\phi_{b}-\phi_{b'})+\cos 2(\phi_{b}+\phi_{b'}-2\,\varphi_b)\,,
\end{eqnarray}
where one has to observe that
\begin{eqnarray}
\label{eqn:interrelation_varphi}
\varphi_b=\varphi_a+\frac{\pi}{2}\,.
\end{eqnarray}
On the right-hand side of Eqs.(\ref{eqn:CorrelFactorFourPhoton_a}) and (\ref{eqn:CorrelFactorFourPhoton_b}) we have 
inserted a factor of 2 for reasons explained in connection with Eq.(\ref{eqn:correlatedProbab}).\\
Since the count rates for all beams are statistically independent once they have passed the beam splitters and the 
polarizing cubes, we have for the total correlation factor
\begin{eqnarray*}
\label{eqn:TotalCorrelFactorFourPhoton}
E_{total}(\varphi_a)=E_{a}(\varphi_a)\,E_{b}(\varphi_b)= \qquad
\end{eqnarray*}
$$
[\cos 2(\phi_{a}-\phi_{a'})+\cos 2(\phi_{a}+\phi_{a'}-2\,\varphi_{b})]\times
$$
$$
[\cos 2(\phi_{b}-\phi_{b'})+\cos 2(\phi_{b}+\phi_{b'}-2 \,\varphi_b)]\,.
$$
This can be recast as
\begin{eqnarray*}
E_{total}(\varphi_a)= \qquad  \qquad \qquad \qquad \qquad \qquad \qquad \qquad \\
\frac{1}{2}\,\{\cos 2(\phi_a -\phi_b -\phi_{a'}+\phi_{b'})+\cos 2(\phi_a +\phi_b -\phi_{a'}-\phi_{b'})
\end{eqnarray*}
$$
+\cos 2[(\phi_a +\phi_{a'} -\phi_{b}-\phi_{b'})-2(\varphi_a -\varphi_b)]
$$
$$
+\cos 2[(\phi_a +\phi_{a'}+\phi_b -\phi_{b'})-2(\varphi_a +\varphi_b)]\}\,.
$$
The first two terms on the right-hand side can be rewritten
$$
\frac{1}{2}\,[\cos 2(\phi_a -\phi_b -\phi_{a'}+\phi_{b'})+\cos 2(\phi_a +\phi_b -\phi_{a'}-\phi_{b'})]=
$$
$$
\cos 2(\phi_a -\phi_{a'})\,\cos 2(\phi_b -\phi_{b'})\,.
$$
If we now make use of Eq.(\ref{eqn:interrelation_varphi}) and average over $\varphi_a$ we arrive at
\begin{eqnarray}
\label{eqn:totalCorrelFactor}
\overline{E}_{total}=\frac{1}{2}\,\cos 2(\phi_a +\phi_{a'} -\phi_{b}-\phi_{b'})+ \nonumber \\
\cos 2(\phi_a -\phi_{a'})\,\cos 2(\phi_b -\phi_{b'})\,.
\end{eqnarray}
The expression $\cos 2(\phi_a +\phi_{a'} -\phi_{b}-\phi_{b'})$ is termed 
``Greenberger-Horne-Zeilinger-(GHZ)-correlation function''.
The factors of the product on the right-hand side have the EPRB-form (\ref{eqn:SpinCorrelation}) and thus refer to 
pairs of ``intra-beam-correlated'' photons. But the product appears only formally as part of the four-photon 
correlation. If the experiment also yields additional true ``intra-beam-correlated'' photons without inter-beam 
correlation that contribution would just appear with a different weight factor in front.\\
Except for such factors in front of the two terms on the right, our result is identical with that obtained by 
Weinfurter and collaborators \cite {Weinfurter}. (It seems, however, that the argument of the $\cos$-functions is 
erroneously by a factor of 2 too small in that article.) The result by Weinfurter and associates is based on a 
completely different reasoning and is thought to provide the ultimate proof that their experiment cannot be 
explained within a local theory. Obviously, that claim, which is also held by almost every researcher in this field, 
is unwarranted.
\section{Conclusions}
Since quantum mechanics is manifestly non-local, it has become a widespread conviction that action at a distance 
constitutes only a feature that reflects this very fact. We have shown that this conclusion is without foundation. 
Effects of non-locality which are most clearly evidenced by two-slit experiments on massive particles, have nothing 
to do with action at a distance, but rather originate in an active role of the vacuum. This has recently been 
demonstrated by the present authors in a  detailed study \cite{Fritsche_Haugk} based on a purely statistical 
(ensemble) interpretation of quantum mechanics which derives from a stochastic foundation. In complete agreement 
with the standpoint taken by Ballentine in his fundamental review article \cite{Ballentine}, we assume that the 
\textit{´´Statistical Interpretation considers a particle to always be at some position in space, each position 
being realized with relative frequency $|\psi(\vec{r})|^2$ in an ensemble of similarly prepared experiments.''}\\
In addition, the particles move along trajectories, which display, however, an irregular departure from their 
classical counterparts. This superimposed quivering motion \textit{(``Zitterbewegung'')} which modifies the 
classical trajectories can sometimes be a negligible effect as with tracks of fast charged particles in a track 
chamber or with trajectories of electrons in an electron field emission microscope. Clearly, individual events like 
the termination of a trajectory on a monitoring screen where it causes the capturing atom to emit a photon, is 
nothing that can be predicted by quantum theory. Again, we are here in complete accord with Ballentine 
\cite{Ballentine} who states: \textit{``...quantum theory is not inconsistent with the supposition that a particle 
has at any instant both a definite position and a definite momentum, although there is a widespread folklore to the 
contrary.''} Furthermore, in support of our approach to photon correlation we refer to another statement in his 
article:
\textit{``Recognition that quantum states should refer to ensembles of similarly prepared systems would seem to open 
the door for hidden variables to control individual events.''}
\\[0.2cm]
As for the unclear terminology ``quantum theory'' (QT) which suggests that there exists a universal theory of 
quantum phenomena comprising ``quantum mechanics'' one should be aware of the fact that there is no complex-valued 
wavefunction in the $3N$-dimensional space that would describe $N$ correlated photons as it does describe $N$ 
correlated electrons, for example.  There is nothing in the quantum mechanics of a massive particle that would 
describe the probability density of finding it at some position in real-space and simultaneously be associated with 
a vector-valued field in that space. Its irregular motion is most clearly reflected in the occurrence of zero-point 
energy when it is bound to an attractive potential. Photons do not perform such an irregular \textit{Zitterbewegung} 
but always move at a constant velocity in vacuo.
\\[0.2cm]
The quantum character of the motion of massive particles is brought out by the modification of their classical 
propagation. In distinct contrast, the propagation of photons is completely controlled by the classical space/time 
behavior of the associated electromagnetic wave. This is most strikingly evi\-denced by the polarizers that are used 
in analyzing the photon correlation experiments. Clearly, these polarizers are designed by exclusively applying 
rules of classical optics. This applies as well to the other optical parts typical of the equipment, viz. mirrors, 
quarter- and halfwave plates, filters and phase shifters.

\end{document}